\documentclass[aps,prl,showpacs,twocolumn,superscriptaddress]{revtex4-1}
\usepackage[pdftex]{graphicx}
\usepackage{dcolumn}
\usepackage{bm}
\usepackage{latexsym}
\usepackage{amssymb}
\usepackage{amsfonts}
\usepackage{amsmath}
\usepackage{color}
\def\be{\begin{equation}}
\def\ee{\end{equation}}

\begin{document}

\title{Iterative Deconvolution of Quadrupole Split NMR Spectra}
\author{Fr\'ed\'eric Mila}
\affiliation{Institute of Theoretical Physics, Ecole Polytechnique F\'ed\'erale de Lausanne (EPFL), CH-1015 Lausanne, Switzerland}
\author{Masashi Takigawa}
\affiliation{Institute for Solid State Physics, University of Tokyo, Kashiwa, Chiba 277-8581, Japan}

\date{\today}

\begin{abstract}
We propose a simple method to deconvolute NMR spectra of quadrupolar nuclei in order to separate the distribution of local magnetic hyperfine field from the quadrupole splitting. It is based on an iterative procedure which allows to express the intensity of a single NMR line directly as a linear combination of the intensities of the total experimental spectrum at a few related frequencies. This procedure is argued to be an interesting complement to Fourier transformation since it can lead to a significant noise reduction in some frequency ranges. This is demonstrated in the case of the $^{11}$B-NMR spectrum in SrCu$_2$(BO$_3$)$_2$ at a field of 31.7 T, where a magnetization plateau at 1/6 of the saturation has been observed.
\end{abstract}

\pacs{75.25.--j, 75.10.Jm, 76.60.Jx}

\maketitle

Over the years, nuclear magnetic resonance (NMR) has proven to be an invaluable tool to identify the local magnetic structure of quantum magnets and strongly correlated electron systems~\cite{Mendels,Takigawa10,Curro,Mukuda}. This relies on the fact that the NMR spectrum is essentially a histogram of the local magnetization, the nuclear frequency being related, via the hyperfine coupling, to the magnetization of the surrounding sites. In particular, one can determine the spin structures and the direction of the moments quite accurately when single crystals are available. 

However, there is one potential complication. If the nuclear spins are larger than 1/2 and the point symmetry of the nuclear site is lower than cubic, each nuclear spin gives rise to more than one line, namely a central line and satellite lines split by the electric quadrupole interaction~\cite{Abragam,Slichter}. The NMR spectrum is then the sum of several histograms, representing the convolution of the distribution of local magnetic hyperfine field and the quadrupole splitting. When the quadrupole splitting between the central and satellite lines is of the same order as the variation of the local hyperfine field, it is necessary to deconvolute the experimental spectrum to obtain the single line spectrum, which directly represents the distribution of local magnetic field. 
 
The authors have recently encountered this problem in studying the $^{11}$B-NMR spectra in the magnetization plateau phases of the quasi-two-dimensional frustrated dimer-spin system SrCu$_2$(BO$_3$)$_2$~\cite{Takigawa13}. This material shows a series of fractional magnetization plateaus, in which complex spin superstructures are formed due to Wigner crystallization of triplet excitations~\cite{Onizuka,Kodama,Takigawa13}. The spectrum in Fig.~1(a) is an example obtained in an external magnetic field of 31.7~T applied along the $c$-axis, where the magnetization shows a plateau at 1/6 of the full saturation. The $^{11}$B-NMR spectra in the plateau phases consist of a central line and two satellite lines, each with an identical line shape corresponding to the superposition of several sharp peaks which originate from unequivalent boron sites coupled to different local magnetizations. Thus, the single line spectrum, once it has been extracted from the original spectrum, can be interpreted as a fingerprint of the spin density distribution in a unit cell of the superstructure.

A standard method to deconvolute the spectra to extract a single line spectrum is based on Fourier transformation (see below). This is straightforward with Fast Fourier Transform (FFT) algorithms. In this paper, we propose a simple alternative to Fourier transformation, which can lead to a significant improvement of the signal to noise ratio of the single line spectrum in certain frequency ranges. It relies on an iterative construction starting either from the low frequency or from the high frequency side of the original spectrum. Note that it is similar in spirit to a method that has been developed a long time ago to resolve the $\alpha_1$ and $\alpha_2$ lines in X-ray diffraction data~\cite{rachinger}. However, to the best of our knowledge, this method has not been extended to the more complicated case of an NMR spectrum with several lines. As we show below, this iterative procedure leads to very simple expressions for the single line spectrum.

Having in mind the example of NMR spectra of $^{11}$B nuclei with nuclear spin 3/2, we start with the case where each nucleus gives rise to one central line and two satellite lines. The general case for an arbitrary size of spin, which is a straightforward generalization, will be briefly discussed at the end. We also assume that the quadrupole splitting is uniform over the entire spectrum, a reasonable assumption if, as in  many magnetically ordered states, the line width due to the inhomogeneous distribution of quadrupole splitting is much smaller than the distribution of magnetic hyperfine field.  

If we denote by $\alpha$ the relative intensity of the central line to its satellites, and by $\nu_Q$ the quadrupole splitting between the central and the satellite lines, the experimental spectrum $g(\nu)$ as a function of the frequency $\nu$ is related to the single line spectrum $f(\nu)$ by
\begin{equation}
g(\nu)=f(\nu-\nu_Q)+\alpha f(\nu)+f(\nu+\nu_Q) .
\label{g_versus_f}
\end{equation}
The theoretical value of $\alpha$ is 4/3~\cite{Abragam}. It would be achieved if the rf-pulse condition was optimized for the central and satellite lines independently. However, this is not a typical situation in experiments, and $\alpha$ should be treated as an adjustable parameter. 
\begin{figure}[t]
\begin{center}
\includegraphics[width=0.9\linewidth]{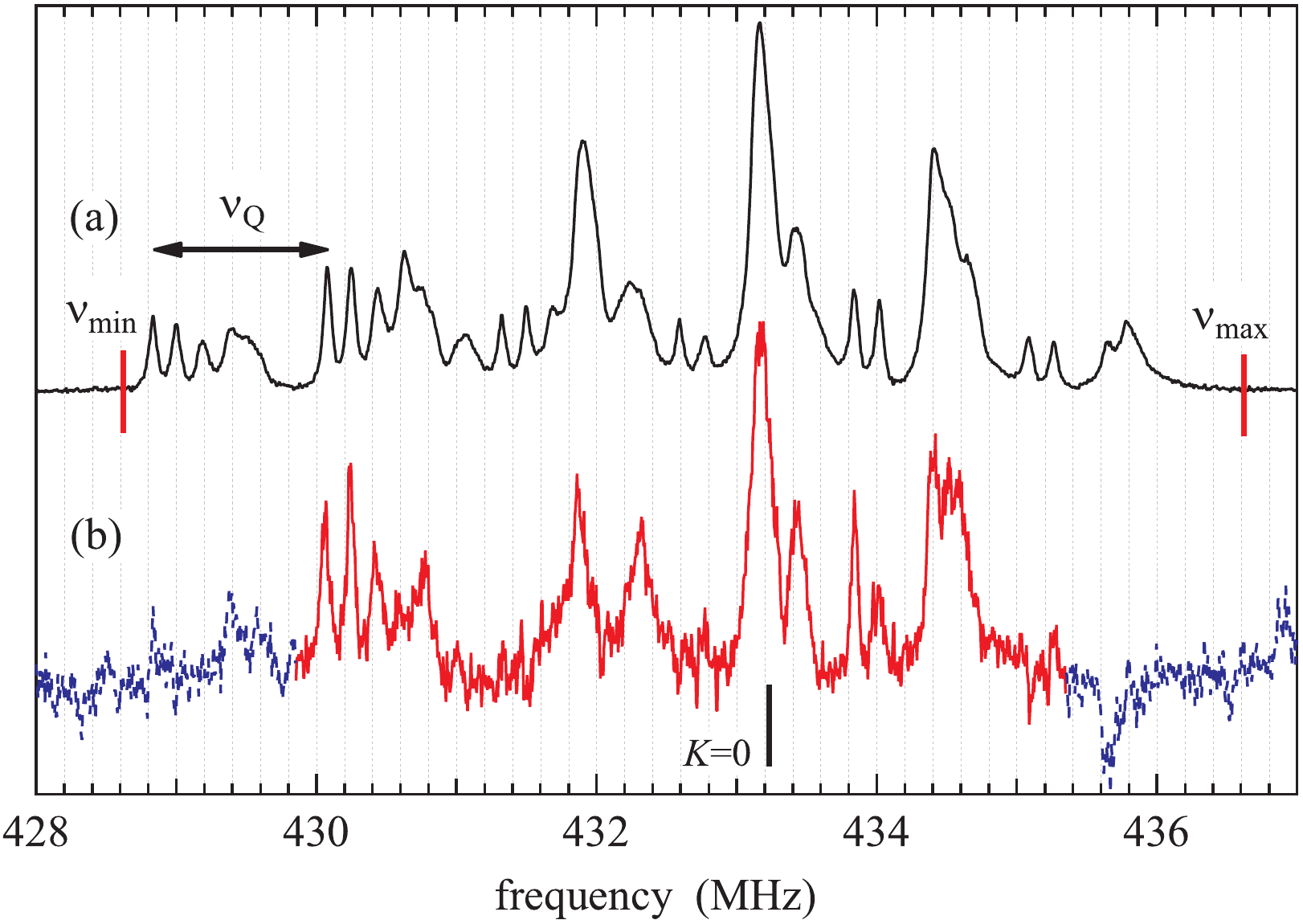}
\end{center}
\caption{(Color online) (a) $^{11}$B-NMR spectrum in the 1/6-plateau phase of SrCu$_2$(BO$_3$)$_2$ obtained at a fixed magnetic field of 31.715~T applied along the $c$-axis with variable frequency. The known value of $\nu_Q$ = 1.25~MHz is shown by the arrow. The spectral intensity is practically zero outside the frequency interval $[\nu_{min},\nu_{max}]$. (b): The deconvoluted single line spectrum obtained from the FFT method with $\alpha$ = 1.75. The deconvoluted spectrum should be zero outside the interval $[\nu_{min}+\nu_Q,\nu_{max}-\nu_Q]$. This physically irrelevant part is shown by the dashed blue lines. The line marked as $K=0$ indicates the NMR frequency in a reference diamagnetic material.}
\label{FFT}
\end{figure}

The problem is to extract $f(\nu)$ from the original $g(\nu)$.
Since Eq.~\ref{g_versus_f} can be expressed as a convolution,
\begin{equation}
g(\nu)=\int_{-\infty}^{+\infty} f(\nu^{\prime}) \phi(\nu-\nu^{\prime}) d\nu^{\prime}
\label{convolution}
\end{equation}
with
\begin{equation}
\phi(\nu)=\delta(\nu-\nu_Q)+\alpha \delta(\nu)+\delta(\nu+\nu_Q), 
\label{Qsplitting}
\end{equation}
a straightforward solution can be obtained by a Fourier transformation. Indeed, taking the Fourier transformation of both
sides of Eq.~\ref{convolution} leads to
\begin{equation}
\tilde g(t)=(\alpha + 2 \cos2\pi \nu_Q t) \tilde f(t)
\end{equation}
where $\tilde g(t)$ and $\tilde f(t)$ are the Fourier transforms of $g(\nu)$ and $f(\nu)$. Then,
with the convention
\begin{equation}
\tilde g(t)=\int_{-\infty}^{+\infty} e^{2\pi i \nu t} g(\nu) d\nu,
\end{equation}
$f(\nu)$ is given by:
\begin{equation}
f(\nu)=\int_{-\infty}^{+\infty} e^{-2\pi i \nu t} \frac{\tilde g(t)}{\alpha + 2 \cos2\pi\nu_Q t} dt.
\label{IFFT}
\end{equation}
This solution is formally very simple, but the result can be rather noisy. The spectrum in Fig.~1(b) is the deconvolution of the experimental spectrum in (a) obtained by this method. The value of 
$\nu_Q$ = 1.25~MHz is known from previous experiments~\cite{Kodama2}.  For the parameter $\alpha$, we have used the value $\alpha$ = 1.75, an estimate obtained by comparing the intensities of the central and satellite lines of specific peaks of the original spectrum. The single line spectrum obtained in this way exhibits a significant amount of noise, even in the frequency range where $f(\nu)$ should be zero (dashed lines in Fig.~1(b), see the discussion preceding Eq.~\ref{condition_on_p} below). Further adjustment of $\alpha$ did not lead to any reduction in the noise. The presence of noise is presumably caused by the zeros of the denominator of Eq.~\ref{IFFT}.

Let us now turn to the iterative solution. For that purpose, we rewrite Eq.{\ref{g_versus_f} as
\begin{equation}
f(\nu)=g(\nu+\nu_Q)-\alpha f(\nu+\nu_Q) - f(\nu+2\nu_Q).
\label{f_versus_g}
\end{equation}
A recursive iteration of this equation, i.e. successive replacements of $f$ in the right hand side by
\begin{eqnarray}
f(\nu+\nu_Q)=g(\nu+2\nu_Q)-\alpha f(\nu+2\nu_Q) - f(\nu+3\nu_Q) , \nonumber \\ 
f(\nu+2\nu_Q)=g(\nu+3\nu_Q)-\alpha f(\nu+3\nu_Q) - f(\nu+4\nu_Q) , \nonumber
\label{iteration}
\end{eqnarray}
and so on, allows one to express $f(\nu)$ as a linear combination of $g(\nu+p \nu_Q)$, $p$ integer:
\begin{equation}
f(\nu)=\sum_{p=1}^{+\infty} c_p \ g(\nu+p\nu_Q).
\label{iteration1}
\end{equation}
This series is in principle infinite, but in practice $g(\nu)$ vanishes outside an interval $[\nu_{min},\nu_{max}]$, and only a finite number of terms are necessary to reconstruct $f(\nu)$. For further reference, the values of $c_p$ up to $p=10$ are listed in Table I. 
The general solution is given in the Appendix.

The order to which one should go is dictated by the interval over which $g(\nu)$ takes significant values. Since peaks of $g(\nu)$ are usually assumed to be lorentzian or gaussian, $g(\nu)$ is strictly speaking never equal to zero, and the series is in principle infinite, but in practice, below the first peak or above the last peak of the full spectrum, $g(\nu)$ rapidly becomes smaller than the noise, which defines a frequency interval $[\nu_{min},\nu_{max}]$ outside which $g(\nu)$ can be considered to be zero. From Eq.\ref{g_versus_f}, it is clear that $f(\nu)$ can itself be considered as zero outside the interval $[\nu_{min}+\nu_Q,\nu_{max}-\nu_Q]$. Then, if $p$ is such that $\nu+p\nu_Q>\nu_{max}$ for $\nu \in [\nu_{min}+\nu_Q,\nu_{max}-\nu_Q]$, the corresponding term is negligible. This implies that all terms with $p>-1+(\nu_{max}-\nu_{min})/\nu_Q$ can be dropped. In other words, in Eq.\ref{iteration1}, it is sufficient to limit the sum to
\begin{equation}
p < \frac{\nu_{max}-\nu_{min}}{\nu_Q}-1.
\label{condition_on_p}
\end{equation}

 \begin{table}
    \begin{displaymath}
      \begin{array}{ccc}
        p & & c_p\\
     & & \\
        1& &1\\
        2& &-\alpha\\
        3& &-1+\alpha^2\\
        4& &2\alpha-\alpha^3\\
        5& &1-3\alpha^2 + \alpha^4\\
        6& &-3 \alpha+4\alpha^3 -\alpha^5\\
        7& \ \ &-1+6 \alpha^2-5 \alpha^4+\alpha^6 \\
       8& & 4 \alpha - 10 \alpha^3 + 6 \alpha^5 - \alpha^7\\
       9& & 1 - 10 \alpha^2 + 15 \alpha^4 - 7 \alpha^6 + \alpha^8\\
       10& & -5 \alpha + 20 \alpha^3 - 21 \alpha^5 + 8 \alpha^7 - \alpha^9\\
       11&\ \  & -1 + 15 \alpha^2 - 35 \alpha^4 + 28 \alpha^6 - 9 \alpha^8 + \alpha^{10}\\
      \end{array}
    \end{displaymath}
    \caption{First 11 coefficients of the series giving $f(\nu)$ as a function of $g(\nu+(p+1)\nu_Q)$.}
  \end{table}

The logic behind this procedure is quite simple. Starting from high frequencies, $f(\nu)=g(\nu+\nu_Q)$ for $\nu_{max}-2\nu_Q \le \nu \le \nu_{max}-\nu_Q$. In that frequency range, $f$ is as precise as $g$. Other contributions enter for smaller values of $\nu$ and one can expect the precision of $f$ to get worse upon lowering the frequency. Indeed, the next peak in $f$ will be mixed in $g$ with other contributions. So this procedure is expected to be more accurate at high 
frequency.

For the low frequency regime, it is a priori better to generate another series along similar lines. The starting point consists in rewriting Eq.\ref{g_versus_f} as 
\begin{equation}
f(\nu)=g(\nu-\nu_Q)-\alpha f(\nu-\nu_Q) - f(\nu-2\nu_Q).
\end{equation} 
Iteration of this equation leads to
\begin{equation}
f(\nu)=\sum_{p=1}^{+\infty} c_p \ g(\nu-p\nu_Q)
\label{iteration2}
\end{equation}
where the coefficients $c_p$ and the condition on $p$ are the same as for the high frequency iteration. Starting from low frequencies, the first peak of $g$ will be directly reflected in $f$, and further peaks in $f$ will be mixed in $g$ with other contributions.

\begin{figure}[t]
\begin{center}
\includegraphics[width=0.9\linewidth]{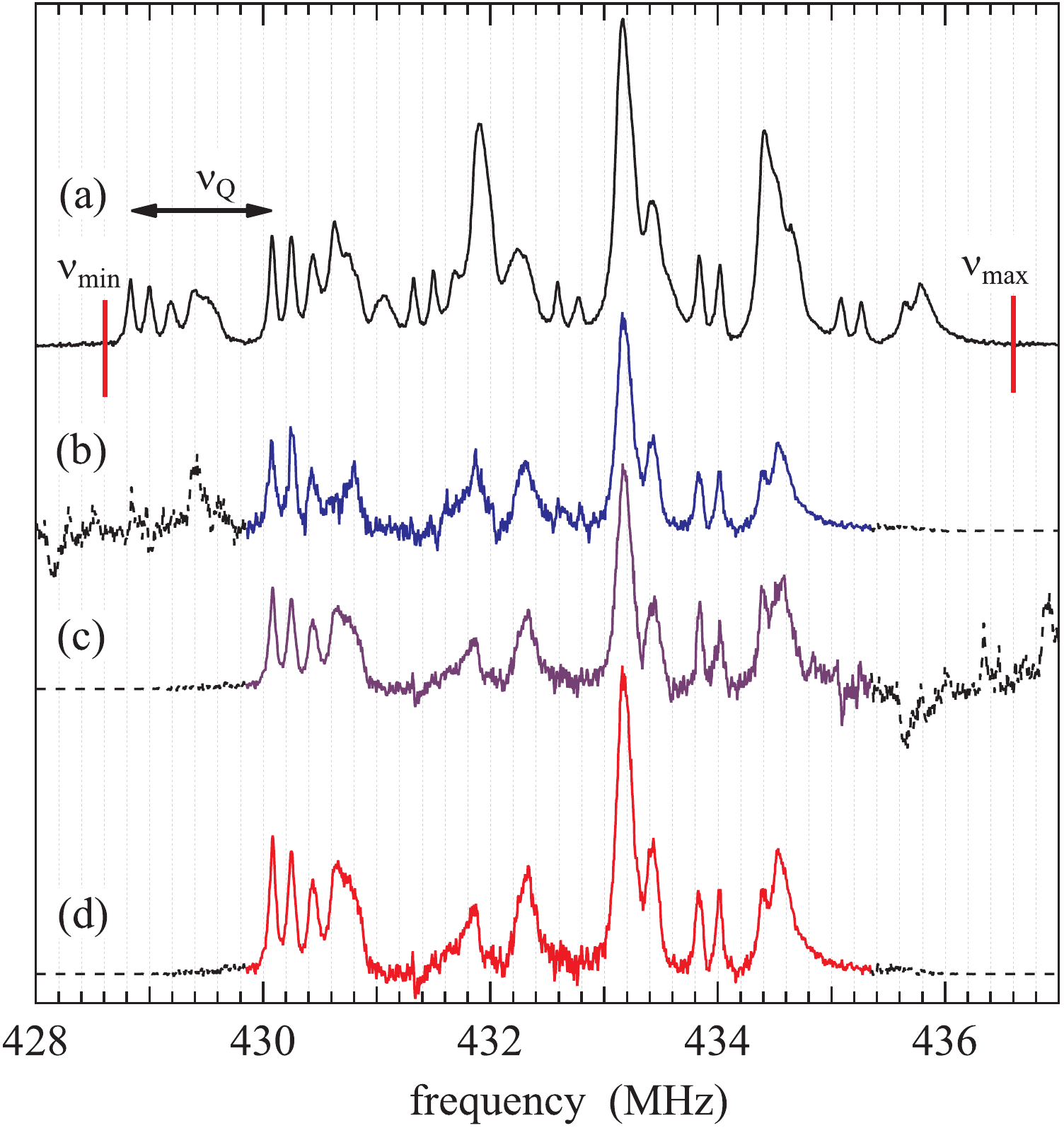}
\end{center}
\caption{(Color online) (a): The same experimental spectrum as in Fig.~1(a). The single line spectrum is obtained by deconvoluting the experimental spectrum from high frequencies using Eq.~\ref{iteration1} (b) or from low frequencies using Eq.~\ref{iteration2} (c). The high frequency part of the spectrum in (b) above 433~MHz is combined with the low frequency part of the spectrum in (c) below 433~MHz to produce the final single line spectrum shown in (d).} 
\label{iteration}
\end{figure}

Let us now test these expressions. We consider again the boron NMR spectrum of SrCu$_2$(BO$_3$)$_2$ at 31.7 T corresponding to the 1/6 plateau.  With $\nu_{min}=428.6$~MHz, $\nu_{max}=436.6$~MHz and $\nu_Q$ = 1.25~MHz, it is only necessary to keep terms up to $p=5$ (see Eq.~\ref{condition_on_p} ).The resulting spectra are shown in Fig.~2. The spectra in (b) and (c) are obtained from the experimental spectrum in (a) by iterative deconvolution from high and low frequencies, respectively, keeping only the first 5 terms in Eq.~\ref{iteration1}. We have used the same value of $\alpha$ = 1.75 as for the Fourier transformation, which turned out to be the optimum value for the smallest spurious signal outside the physically relevant frequency range. The noise level depends on the frequency range and varies significantly from one method to the other: it is very low at low frequency if one starts from low frequencies, and very low at high frequency if one starts from high frequencies. In the present case, a simple inspection of Figs.~2(b-c) shows that the noise
in Fig.~2(b) is larger than that of Fig.~2(c) below approximately 433 ~MHz and smaller above. The final result shown in Fig.~2(d) has been obtained by taking the high frequency part of the spectrum in (b) above 433~MHz and
the low frequency part of the spectrum in (c) below 433~MHz, and by pasting them together. The resulting spectrum can be considered
as a very reliable single line spectrum.

The results are similar to those obtained by Fourier transformation and shown in Fig.~1, as they should of course since, in the absence of experimental noise, they would be strictly identical. However, the iterative deconvolution has two main advantages. First of all, it is able to give a single line spectrum that has very small noise both at high and at low frequencies. As a consequence, features in the intermediate frequency range  that consistently appear when starting from either low or high frequencies can be considered as reliable features of the single line spectrum. Regarding the boron spectrum of SrCu$_2$(BO$_3$)$_2$ at 31.7 T, this iterative deconvolution procedure has been instrumental in identifying the actual sequence of peaks, and in getting a good estimate of their intensity. In addition, the iterative deconvolution is very simple: the single line spectrum at a given frequency is just a linear combination of the values of the original spectrum at a few frequencies.

Finally, let us briefly comment on the more general case for larger nuclear spin, $I > 3/2$, leading to $2I$ quadrupole split lines. The full spectrum is a linear combination of $2I$ single line spectra, and an equation similar to Eq.~\ref{f_versus_g} can be written, in which the function $f$ will appear $2I-1$ times in the right hand side at frequencies $\nu+\nu_Q$, $\nu+2\nu_Q$,... $\nu+(2I-1)\nu_Q$.  Iterating this equation will lead again to a general relation between $f$ and $g$ of the form of Eq.~\ref{iteration1}, but with different coefficients that depend on $I$ and on the relative intensities of the satellites. These coefficients can be obtained by recursive iteration, or by reformulating the problem as a recurrence of step $2I-1$ (see Appendix).

In conclusion, a simple alternative to Fourier Transform for the deconvolution of NMR spectra has been worked out. It allows to reach a significantly better signal-to-noise ratio. This new way of performing the deconvolution has played an important role in interpreting the high field NMR results in SrCu$_2$(BO$_3$)$_2$. It is our hope that it will prove useful in other contexts as well.

We would like to express our sincere thanks to C. Berthier and M. Horvati\'c for invaluable discussions on the NMR spectra of SrCu$_2$(BO$_3$)$_2$ and very useful comments on the manuscript. This work has been performed while one of the authors, F. M., was a visiting professor at ISSP, whose hospitality is gratefully acknowledged.

\section{Appendix}

To get an explicit form of the coefficients of this series, we note that Eq.\ref{f_versus_g} defines a recurrence of step 2. We thus rewrite it in matrix notation:
\begin{equation}
\left(\begin{array}{c} f(\nu) \\ f(\nu+\nu_Q) \end{array}\right)=\left(\begin{array}{c} g(\nu+\nu_Q) \\ 0\end{array}\right)+A \left(\begin{array}{c} f(\nu+\nu_Q) \\ f(\nu+2\nu_Q) \end{array}\right)
\end{equation}
where $A$ is a $(2 \times 2)$ matrix defined by
\begin{equation}
A=\left(\begin{array}{cc}
			      -\alpha & -1 \\
			      1 & 0
			      \end{array}\right)
\end{equation}
Iterating leads to
\begin{equation}
\left(\begin{array}{c} f(\nu) \\ f(\nu+\nu_Q) \end{array}\right)=\sum_{p=1}^{+\infty} A^{p-1} \left(\begin{array}{c} g(\nu+p\nu_Q) \\ 0 \end{array}\right)
\end{equation}
with the convention that $A^0$ is the identity. The upper component of this equation is the expression we
are looking for. It reads:
\begin{equation}
f(\nu)=\sum_{p=1}^{+\infty} (A^{p-1})_{11} \ g(\nu+p\nu_Q)
\label{main_result}
\end{equation}
In other words, the coefficients of Eq.\ref{iteration1} are given by $c_p=(A^{p-1})_{11}$, an expression from which 
coefficients up to arbitrary high order can be easily obtained. 

Although this is of limited practical
use, we note for completeness that an explicit expression can be obtained for the coefficients $c_p$ by diagonalizing the matrix $A$:
\begin{equation}
c_p=\frac{(-\alpha+\sqrt{-4+\alpha^2})^p-(-\alpha-\sqrt{-4+\alpha^2})^p}{2^p\sqrt{-4+\alpha^2}}
\label{solution}
\end{equation}
Since only even powers of the square roots appear in the right hand side of Eq.~\ref{solution}, one does not need to worry about whether $\alpha$ is larger or smaller than 2. In fact, $c_p$ can be reexpressed as a polynomial of degree $p$ in $\alpha$:
\begin{equation}
c_p=\frac{1}{2^p} \sum_{\begin{array} {c} n=1 \\ n\ {\rm odd}\end{array}}^p\left(\begin{array}{c} p \\ n \end{array}\right) (-\alpha)^{p-n} (-4+\alpha^2)^\frac{n-1}{2}
\end{equation}ее
where $\left(\begin{array}{c} p \\ n \end{array}\right)$ stands for the binomial coefficient.

In the general case of $2I$ quadrupole split lines, the coefficients $c_p$ are still given by $(A^{p-1})_{11}$, but with a matrix $A$ of dimension $(2I-1) \times (2I-1)$ whose coefficients are simply determined by expressing the relation between $f$ and $g$ as a recurrence of step $2I-1$.

\end{document}